\documentclass[showpacs,pra,showkeys,twocolumn,groupedaddress,superscriptaddress,amsmath,amssymb]{revtex4}
\usepackage{graphicx}
\usepackage{dcolumn}
\usepackage{bm}

\begin{document}

\title{The Strange Man in Random Networks of Automata}



\author{Carlos Handrey A. Ferraz}
\email{handrey@fisica.ufc.br}
\affiliation{Physics Departament, Universidade Federal do Ceara-UFC}

\author{Hans J. Herrmann}
\email{hans@fisica.ufc.br}
\affiliation{Physics Departament, Universidade Federal do Ceara-UFC}
\affiliation{Computational Physics, IfB, ETH Z\"urich H\"onggerberg 8098 Z\"urich Switzerland}


\date{\today}

\begin{abstract}
 We have performed computer simulations of Kauffman's automata on several graphs  such
as the regular square lattice and invasion percolation clusters in order to investigate phase transitions, radial distributions of the mean total damage (dynamical exponent $z$) and propagation speeds of the damage when one adds a damaging agent, nicknamed "strange man".
Despite the increase in the damaging efficiency,  we have not observed any appreciable change at the transition threshold to chaos neither for the short-range nor for the small-world case on the square lattices when the strange man is added in comparison to when small initial damages are inserted in the system. The propagation speed of the damage cloud until touching the border of the system in both cases obeys a power law with a critical exponent $\alpha$ that strongly depends on the lattice. Particularly, we have ckecked the damage spreading when some connections are removed on the square lattice and when one considers special invasion percolation clusters (high boundary-saturation clusters). It is seen that the propagation speed in these systems is quite sensible to the degree of dilution.\\

\end{abstract}


\maketitle

\section{Introduction \label{int}}
The study of cellular automata  has interested physicists  because of 
their many applications \cite{wolfran1,stauf1} and their contributions to theoretical physics \cite{vich1}. In fact, problems of cellular automata are very related to classical models of statistical
mechanics (percolation, spin-glass, Ising model). Furthermore cellular automata can 
be seen as an alternative to equations of motion \cite{vich2}, although, nobody has yet found 
a clear way to switch from one description to the other. A variety of striking phenomena as 
stability, criticality, fractals and chaos, can be discussed within the technique of automata.
In particular, we are interested in studying how small failures (damages) produced 
at complex structures of automata propagate throughout the system.  A biological interpretation can be given to the problem of mutations on biological genes. A mutation could be a change at any particular point (site) of the genetic material which affects the 
time development of the system. The diference between the genetical material with mutation and without occurrence of mutation is described by the Hamming distance of the system. 
Other authors \cite{derr,stauf2,arcang,stauf3} have made computer simulations finding a phase 
transition as well as fractal behaviour in various structures by inserting a small initial damage and then allowing the system to evolve in time. But what does happen if one introduces an agent that damages the system all the time? 
We will call this agent "strange man" because his rules are randomly updated during the whole process, i.e., he is a probabilistic automaton among deterministic automata.  
We gave him this nickname in analogy to a social model where each individual follows the rules very well, except one. We have performed numerical simulations in order to investigate phase transitions, radial distributions of the mean total damage and propagation speeds of the damage under such conditions. We have dealt with well-known networks of automata such as regular square lattices (short-range case), invasion percolation clusters \cite{wilk} and also the small-world topology (small-world case)\cite{watt} by making use of one of the most important and physically meaningful automata model, namely the Kauffman model. The Kauffman model or more generally random Boolean networks have got renewed interest because of their recent applications to sychronization \cite{greil}, stability \cite{bilke}, control of chaos \cite{luque} and scale-free topologies \cite{aldana,iguchi}. In our simulations, despite of the increase in the damaging efficiency, we have not observed any appreciable change of the transition threshold to chaos neither for the short-range nor for the small-world case on the square lattice when the strange man is added in comparison to when only small initial damages are inserted in the system. Nevertheless, the propagation speed of the damage cloud until touching the system boundaries  in both cases is substantially increased and obeys a power-law. For the short-range case, we have used an extrapolation technique in order to reduce finite-size effects on the square lattice. Furthermore, we have also checked the damage spreading when some connections are removed from the square lattice and when one considers invasion percolation clusters with high boundary-saturation. It is seen that the propagation speed in these systems is quite sensitive with respect to dilution. In particular, the  damage remains confined around the strange man when only few connections are removed (typically less than 20\%). That shows the existence of unstable zones \cite{bastolla} in the system, i.e., zones formed by susceptible sites over which the damage can spread. 

\section{The Kauffman Model}
In the past, Kauffman \cite{kauff} introduced networks of Boolean automata in order to study the behaviour of generic regulatory systems. The basic idea of the 
Kauffman model is to consider a mixture of all possible binary cellular automata. The Kauffman 
model can be realized on a lattice, where Boolean rules are chosen separately for each 
site.  Each of $N$ lattice sites hosts a Boolean variable  $\sigma_{i}$ ( spin up or down) which
is either zero or unity. The time evolution of this model is determined by $N$ functions $f_{i}$
(rules) which are randomly chosen for each site independently, and by the choice of $K$ input
sites \{$j_{K}(i)$\} for each site $i$. Thus the value $\sigma_{i}$ 
at site $i$  for time $t$+1 is given by:
\begin{equation}\label{eq:1}
\sigma_{i}(t+1)=f_{i}(\sigma_{j_{1}}(t), \ldots,\sigma_{j_{K}}(t))\qquad i=1,2,\ldots,N.
\end{equation} 
Each Boolean function $f_{i}$ is specified, once its value is given for each of the $2^{K}$ possible 
neighbour configurations. A variable $\sigma_{i}$ is said relevant for the spreading damage process if it is unstable and/or the state of  other variables \{$\sigma_{j}$\} depend on $\sigma_{i}$.  Moreover if one imposes that both the inputs and the chosen Boolean functions do not change with the time, we have the quenched Kauffman model. On the other hand, if one admits that both change with time, we have the annealed Kauffman model. 
A big difference between the two cases is that in the quenched case there are limit cycles and in the annealed case not. As the time development in this case is totally deterministic, and since $N$ different Boolean variables can produce $2^{N}$ different lattice configurations, we must return after at most $2^{N}$ time-steps to the previous initial configuration. Then the system will repeat the same  configurations, without ever leaving this limit cycle. For the nearest-neighbour Kauffman model on the square lattice, the number of relevant limit cycles increases exponentially with system size in the non-chaotic phase, as shown by Derrida and Flyvbjerg \cite {derr2}. Kauffman identified these different limit cycles with the different cell types in our body and found that their number grows as $\sqrt{N}$ for $N$ interacting genes. Thus the nearest-neighbour square lattice is not the most realistic model for such biological aspects. The annealed case can be solved analytically, whereas for the quenched case only computer simulations were performed up to now. 

\section{Computational Procedure}
A standard fashion to implement the Kauffman model on computers is introducing a parameter $p$ such that for each site on the lattice we select among the ${{2}^2}^K$ rules one which for each outcome will have  spin up with probability $p$. In a computer simulation, first one goes through all $N$ sites of the system, and for each site one goes through all ${2}^K$ neighbour configurations, and for each such configuration one determines by drawing one random number if its spin will be up or down; if the random number is smaller than $p$ then its spin will be up, otherwise it will be down. Once one has gone through all neighbour configurations of that site, then one has fixed the rule for that site, and one can go to the next site. After that one selects an initial configuration of the Boolean variables by randomly assigning to each lattice site a spin up or down with equal probability.  We will consider two systems (replicas), identical in the connections and rules, and also identical in the initial configuration of the Boolean variables, except that on one of them we put the strange man on a central site of the lattice.  The number of spins which at time $t$ are different between the two replicas is called the Hamming distance $d(t)$  or simply the damage. For two lattice configurations \{$\sigma_{i}(t)$\} and \{$\rho_{i}(t)$\}, we have
\begin{equation}\label{eq:2}
d(t)=\dfrac{1}{N}\sum_{i}|\sigma_{i}(t)-\rho_{i}(t)|,
\end{equation}
and we can define an order parameter $\psi$ for the system taking in Eq. \ref{eq:2} the limit $t\rightarrow \infty$ , that is
\begin{equation}
\psi=\lim_{\substack{d(0)\rightarrow 0}}d(\infty).
\end{equation}
Computationally, convergence is typically reached after some few thousand  time steps. In this way we can study both the phase transition and the propagation speed of the damage cloud on all networks treated here, by varying the value $p$. Obviously, $p$ and $1-p$ are statistically equivalent, so that we do not consider $p>0.50$. In order to take into account the small-world case, we have also introduced a rewiring probability $q$ which rewires the $K=4$ inputs of each site on the square lattice to any other site, not necessarily being its nearest-neighbour. For more details about this rewiring procedure as well as a phase transiton and fractality in the Kauffman model on small-world topology, the reader should consult Ref.\cite{hand}.
 \begin{figure*}[t]
\begin{minipage} [!l]{0.48\linewidth}
\includegraphics*[scale=0.40,angle=-90]{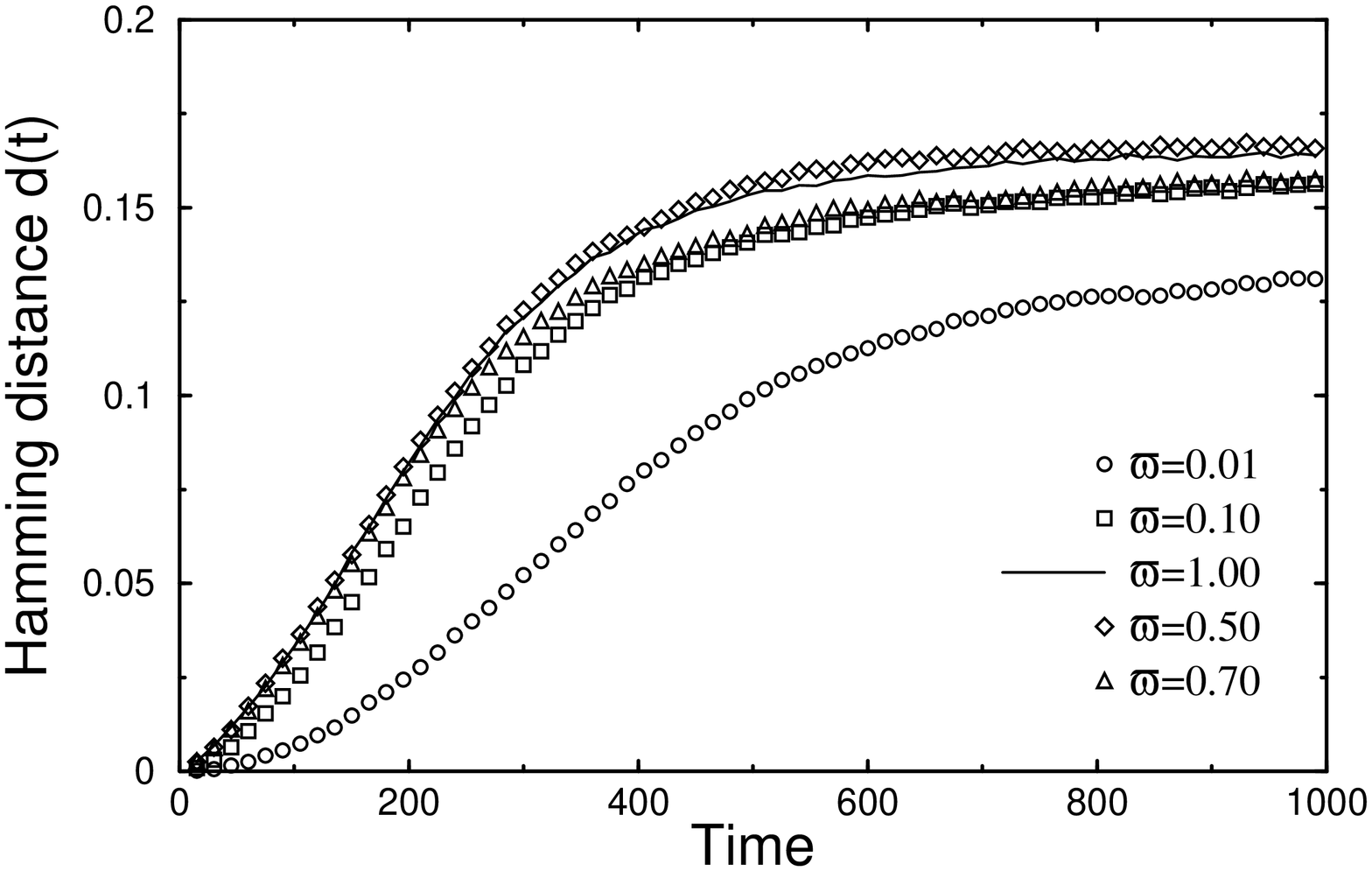}
\caption{Time evolution of the Hamming distance on a 80$\times$80 lattice  for different values of $\varpi$ at $p=0.30$.}\label{fig1}
\end{minipage}\hfill
\begin{minipage}[!l]{0.48\linewidth}
\includegraphics*[scale=0.40,angle=-90]{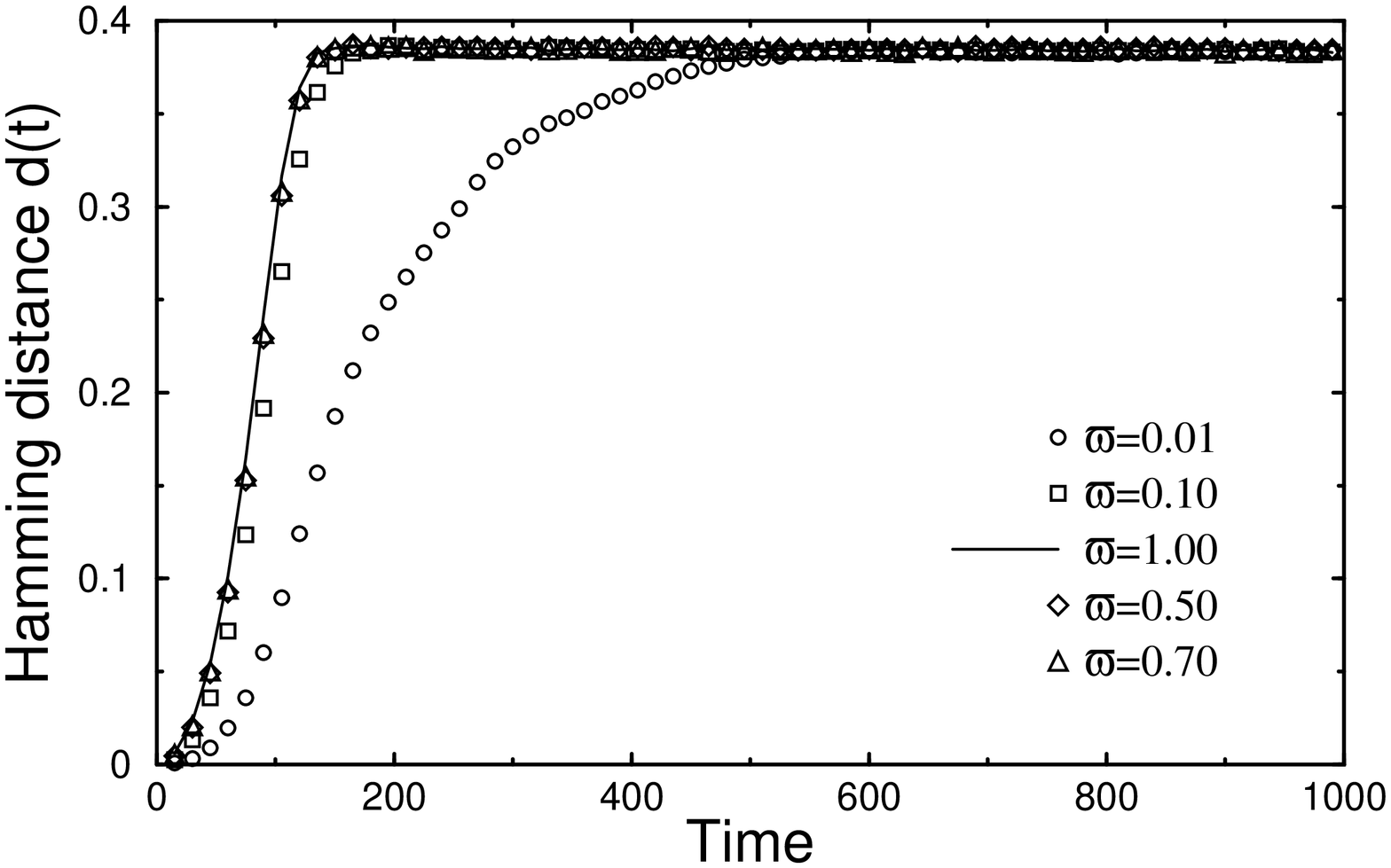}
\caption{Time evolution of the Hamming distance on a 80$\times$80 lattice for different values of $\varpi$ at $p=0.40$.}\label{fig2}
\end{minipage}
\end{figure*} 
In the present work, we have regarded both the quenched and the annealed version. It is worth noting that for the annealed version in the infinite range limit, Eq. \ref{eq:2} can be iterated \cite{derr3} as
\begin{equation}
d(t+1)= 2p(1-p)(1-(1-d(t))^{K}),
\end{equation} 
where $(1-d(t))^{K}$ is the fraction of sites that at the next time-step have the same $K$ inputs. 
 
Invasion percolation clusters were created starting from a square lattice and assigning to each lattice site a random number between 0 and 1. The algorithm to produce such a cluster is as follows: First one chooses a central site on the lattice to be the initial seed of the cluster. Then one chooses the site with the smallest random number among its nearest-neighbours and adds this site to the cluster. That defines a new set of nearest-neighbours for the cluster, i.e., a new boundary of the cluster. Then again one connects to the cluster the site with the smallest random number belonging to this new boundary and so on. The growth process continues until the cluster touches the  borders of the square lattice a certain number of times. This multiple touching on the lattice borders has the effect of increasing the number of connections in the cluster inside the square lattice. Such a procedure was necessary to guarantee the damage spreading throughout the system. We called the cluster built with this procedure \ "high boundary-saturation cluster" (HBSC). In our algorithm, the number of touching sites on the lattice borders  is controlled by the variable $\xi$ (boundary-saturation). If $\xi=0.80$, then 80\% of the sites of the border of square lattice will be part of this HBSC.  Once the HBSC has been created, the strange man is placed at the initial seed of one of the cluster replicas and the comparision between replicas is then carried out. 

Furthermore we have considered a diluted square lattice Kauffman model. The inputs of nearest-neighbours now vary from  $K=1$ to  $K=4$ for each site. That was achieved by introducing a connection probability $p_{link}$ for each of the four inputs of the sites and imposing at least one input per site. The sites $i$ are treated one by one. One goes clockwise drawing a random number for each one of its four inputs \{$\sigma_{j}$\}; if the random number is smaller than $p_{link}$ then that input $\sigma_{j}$ is a variable of the Boolean function $f_{i}$ [Eq. \ref{eq:1}], otherwise it is not. This looping procedure is repeated  until at least one input is a variable of the Boolean function. In this way disconnected sites of the lattice are avoided in our simulations. 
 
We also define another probability called the strange man's behaviour pattern $\varpi$ controlling how many times the strange man damages the system per time unit. It could be understood as its activity rate. So if one regards a strange man with $\varpi=0.50$ then for each time step he has a 50\% chance to randomly change his rule.

The quenched version was studied both on the HBSC and on the square lattice. We have used the annealed version only on invasion percolation clusters to calculate both the propagation speed of the damage cloud and the radial distributions of the mean total damage on these structures. In this case the Boolean functions change after each time step.

\section{Numerical Results}
As a preliminary test for about how the strange man acts in an automata system, we simulate two replicas of a square lattice in order to analyze the temporal evolution of the Hamming distance up to 1000 time-steps for strange men with different behaviour patterns $\varpi$.  In Figs.\ref{fig1} and \ref{fig2}, we show the Hamming distance on a square lattice with 6400 sites and five different values of $\varpi$ at $p=0.30$ and $p=0.40$ respectively, averaged over 100 different initial configurations. As we can see from Figs. \ref{fig1} and \ref{fig2}, changes in the values of $\varpi$ are only relevant in the transient state. Even for a rather small value for $\varpi$ as 0.01 the system converges rapidly towards the stationary state. Due to the little impact caused by changing $\varpi$ when one considers the asymptotic time development of the system,  we will from now on always keep $\varpi=1.0$ along this paper. However a variation in $\varpi$ could imply a significant change in the touching time on the lattice boundaries, since this happens in the transient state.
 
\begin{figure*}[t]
\begin{minipage} [!l]{0.48\linewidth}
\includegraphics*[scale=0.40,angle=-90]{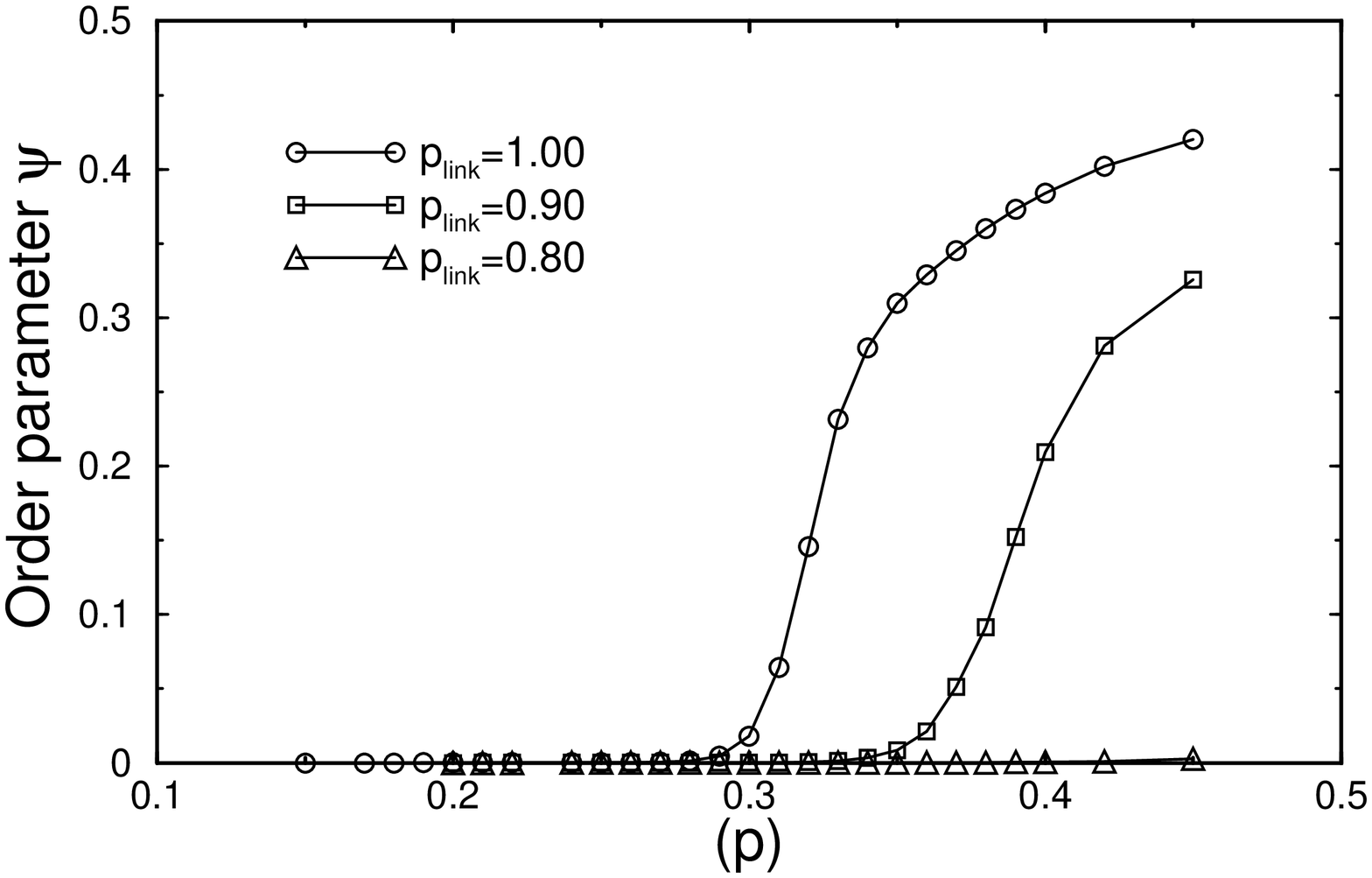}
\caption{Phase transition on a 500$\times$500 lattice for three different values of the probabilities $p_{link}$ The critical point at $p_{link}=1.0$ is 0.31, at $p_{link}=0.90$ is 0.37 and at $p_{link}=0.80$ one no longer has a phase transition.}\label{fig3}
\end{minipage}\hfill
\begin{minipage}[!l]{0.48\linewidth}
\includegraphics*[scale=0.40,angle=-90]{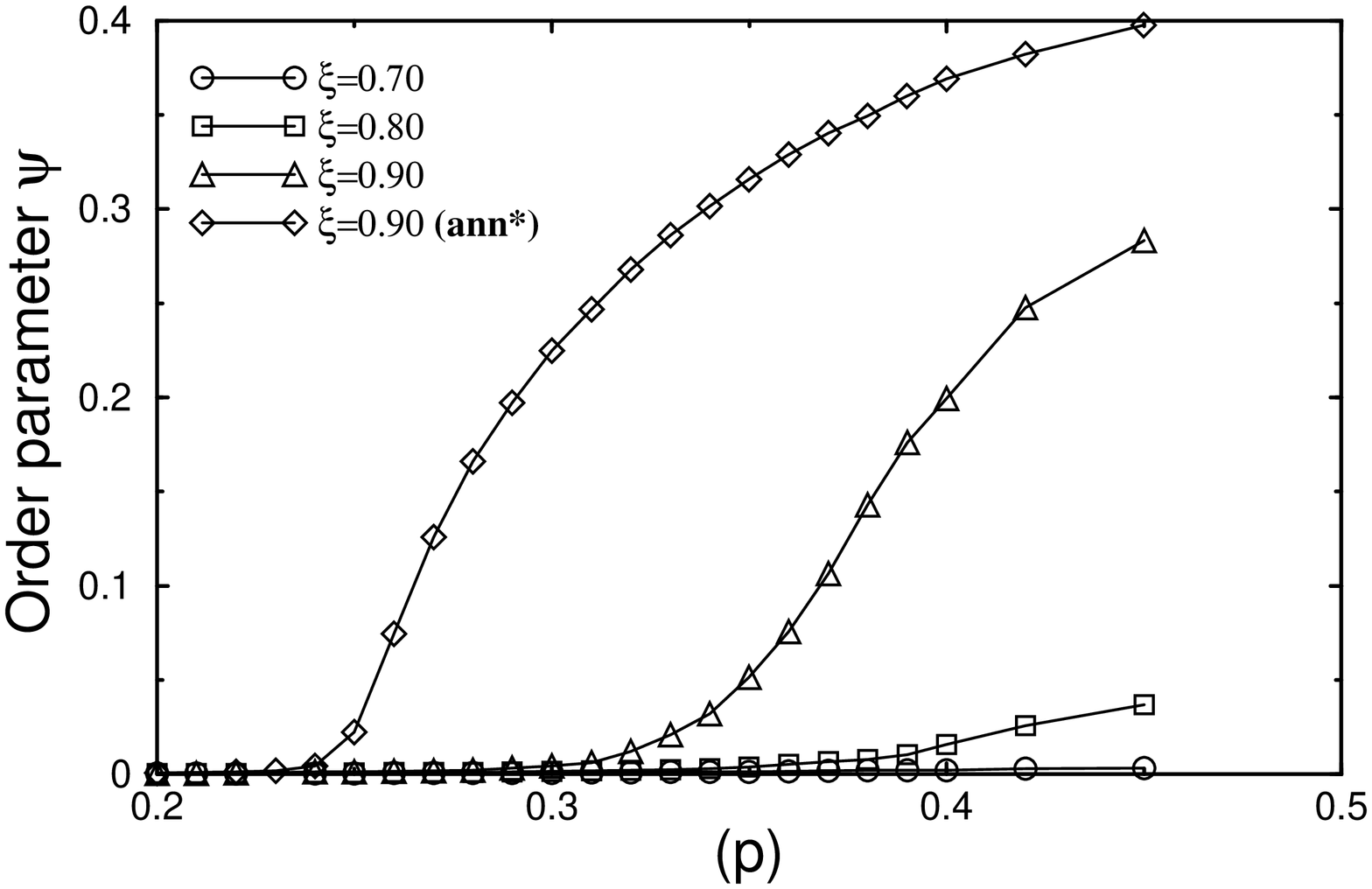}
\caption{Phase transition on a HBSC built on a 100$\times$100 lattice for 
three different values of the variable $\xi$ in the quenched case. Particularly, at $\xi=0.90$ we show our result for the annealed case as well. At $\xi=0.90$ one has $p_{c}=0.26$ (annealed case) and $p_{c}=0.36$ (quenched case).}\label{fig4}
\end{minipage}
\end{figure*} 
\begin{figure}[h]
\includegraphics*[scale=0.40,angle=-90]{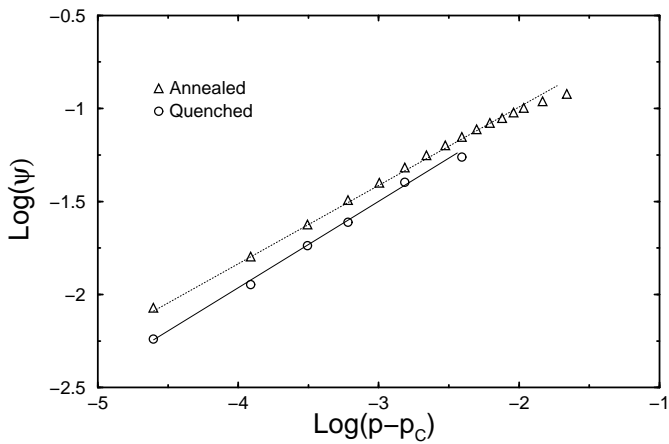}
\caption{Log-log plot of $\psi$ vs $(p-p_{c})$ for a HBSC built on a 100$\times$100 lattice for both  the annealed and the quenched version.}\label{fig4a}
\end{figure} 

Another interesting issue is to search for a phase transition on the square lattice when one introduces a strange man in the system. This is done for a 500$\times$500 lattice by calculating the order parameter $\psi$ averaged over 100 runs with up to 1000 time-steps, where we have also studied the impact caused when one cuts randomly some of their connections, as shown in Fig. \ref{fig3}. Two conclusions can be drawn: On one hand, the continuous activity of the strange man does not affect the value of the critical point which for the square lattice is at 0.31 according to Ref. \cite{stauf3}. 
The same thing also happens if one regards the small-world case \cite{hand}. On the other hand, the damage spreading  on square lattices is quite sensitive to the removal of connections. That is due to the existence of unstable regions on the lattice (unstable cores). As we can see from Fig. 3, taking a connection probability $p_{link}=0.90$, the value of the critical point on the lattice is pushed to $p=0.37$. Yet more dramatic, at $p_{link}=0.80$ we do not find a chaotic phase any more.  

 \begin{figure*}[t]
\begin{minipage} [!l]{0.48\linewidth}
\includegraphics*[scale=0.40,angle=-90]{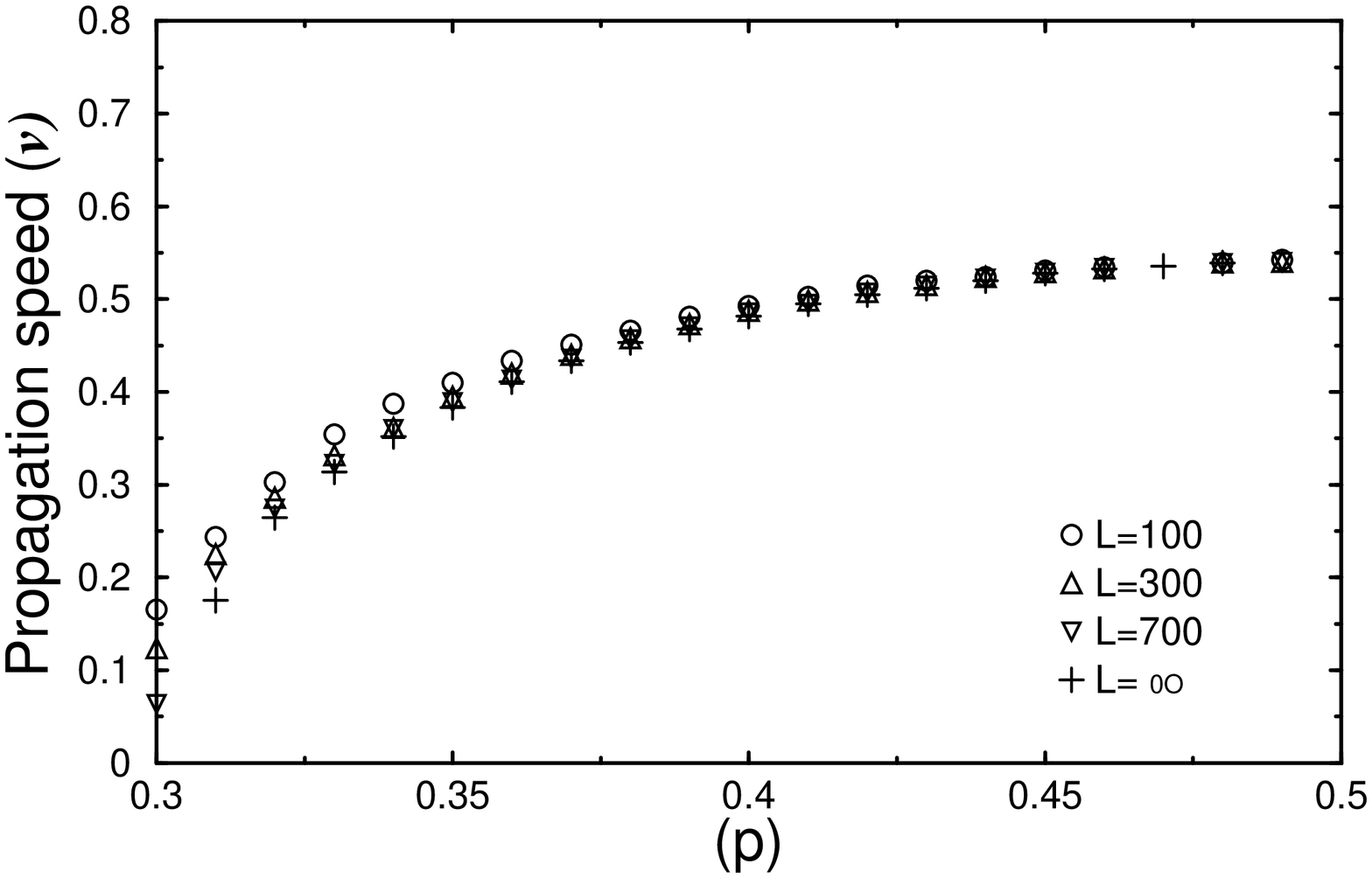}
\caption{Propagation speed of the damage in the short-range case ($q=0.0$), for lattice sizes $L$=100, 300, 700 along with an extrapolation of these data ($L\rightarrow\infty$). }\label{fig5}
\end{minipage}\hfill
 \begin{minipage}[!l]{0.48\linewidth}
 \includegraphics*[scale=0.40,angle=-90]{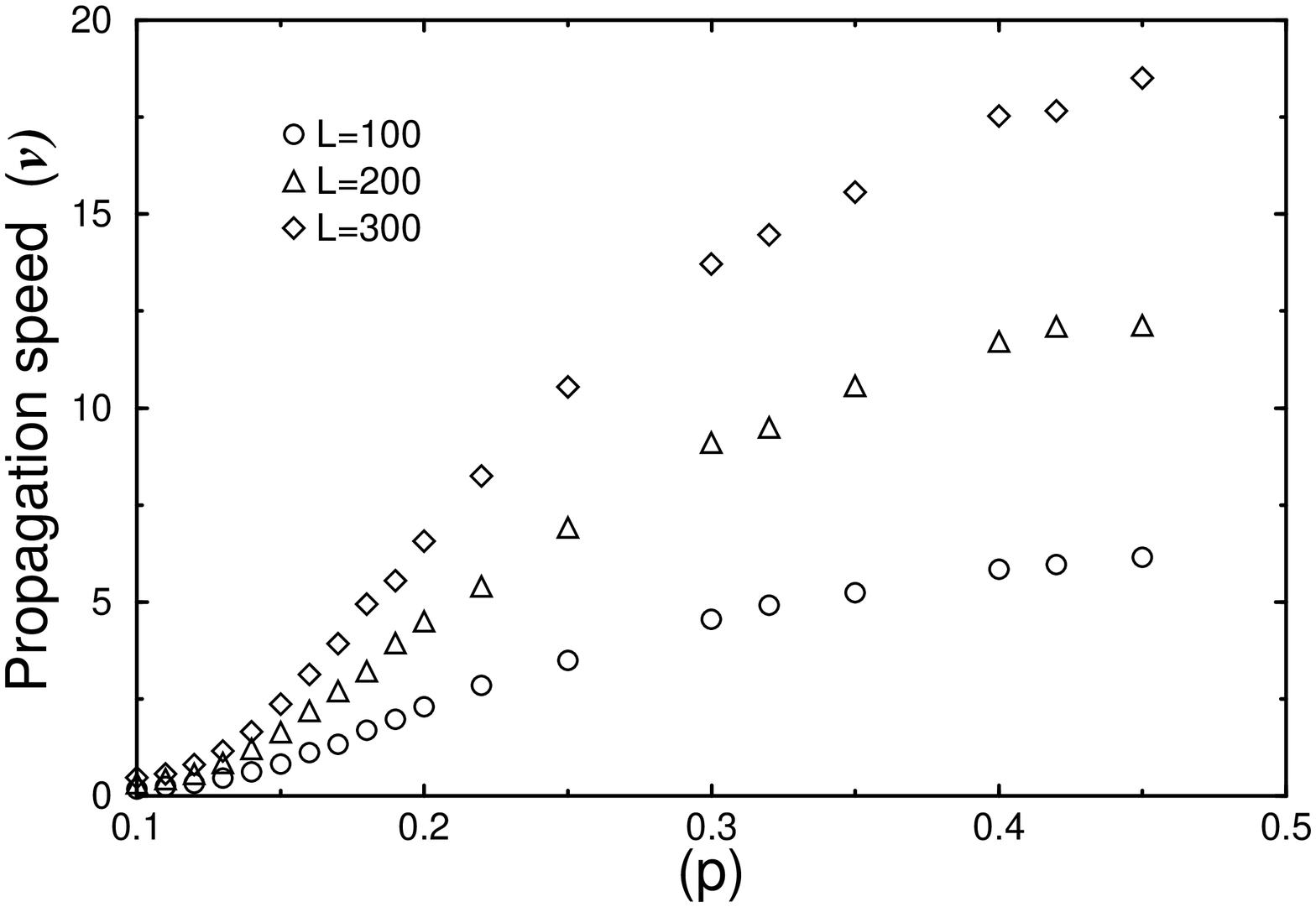}
 \caption{Propagation speed of the damage in the small-world case ($q=0.05$), for lattice sizes 
  $L$=100, 200 and 300.}\label{fig6}
 \end{minipage}
\end{figure*} 

We have done the same analysis for the order parameter $\psi$  on a HBSC built on a 100$\times$100 lattice, where $N$ in Eq. \ref{eq:2} is the number of sites of a typical cluster. Our results are shown in Fig. \ref{fig4} for three different values of $\xi$ (see previous section). Particularly, at $\xi=0.90$ we also present a simulation for the annealed version on this cluster. For a boundary-saturation $\xi=0.90$, we can see a phase transition around $p_{c}=0.36$ in the quenched version and around $p_{c}=0.26$ in the annealed version. In Fig. \ref{fig4a} we show a  log-log plot of $\psi$ vs $(p-p_{c})$. Our estimates for the critical exponents $\beta$ for both versions are:
\begin{align}
\beta^{\text{ annealed}}&=0.42\pm 0.02 \\ \beta^{\text{ quenched}}&=0.47\pm 0.02
\end{align}
\begin{figure}[h]
\includegraphics*[scale=0.40,angle=-90]{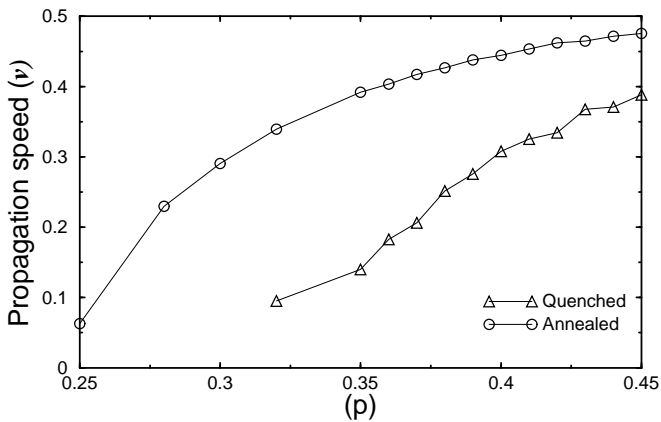} 
\caption{Propagation speed of the damage on the HBSC at $\xi=0.90$ inside a 100$\times$100 lattice both in the quenched and the annealed case.}\label{f6}
\end{figure} 
\begin{figure}[h]
\includegraphics*[scale=0.40,angle=-90]{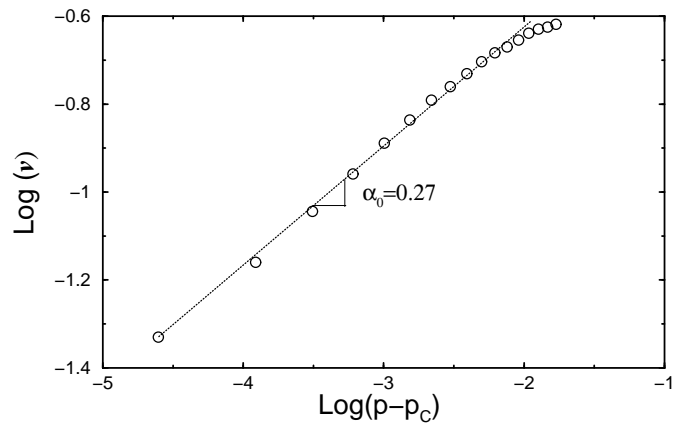}
\caption{Log-log plot of $v$ vs $(p-p_{c})$ for the extrapolated data from Fig. \ref {fig5} in the  short-range case ($q=0.0$).}\label{fig7}
\end{figure} 
\begin{figure}[h]
\includegraphics*[scale=0.40,angle=-90]{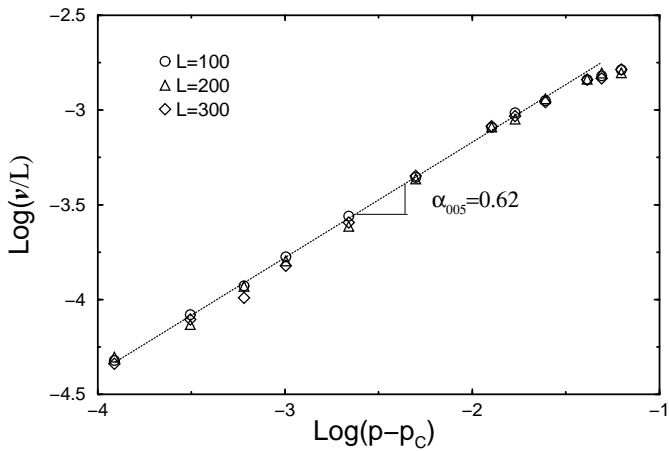}
\caption{Log-log plot of $v/L$ vs $(p-p_{c})$ of the data plotted in Fig. \ref{fig6} in the small-world case ($q=0.05$).}\label{fig8}
\end{figure} 
\begin{figure}[h]
\includegraphics*[scale=0.40,angle=-90]{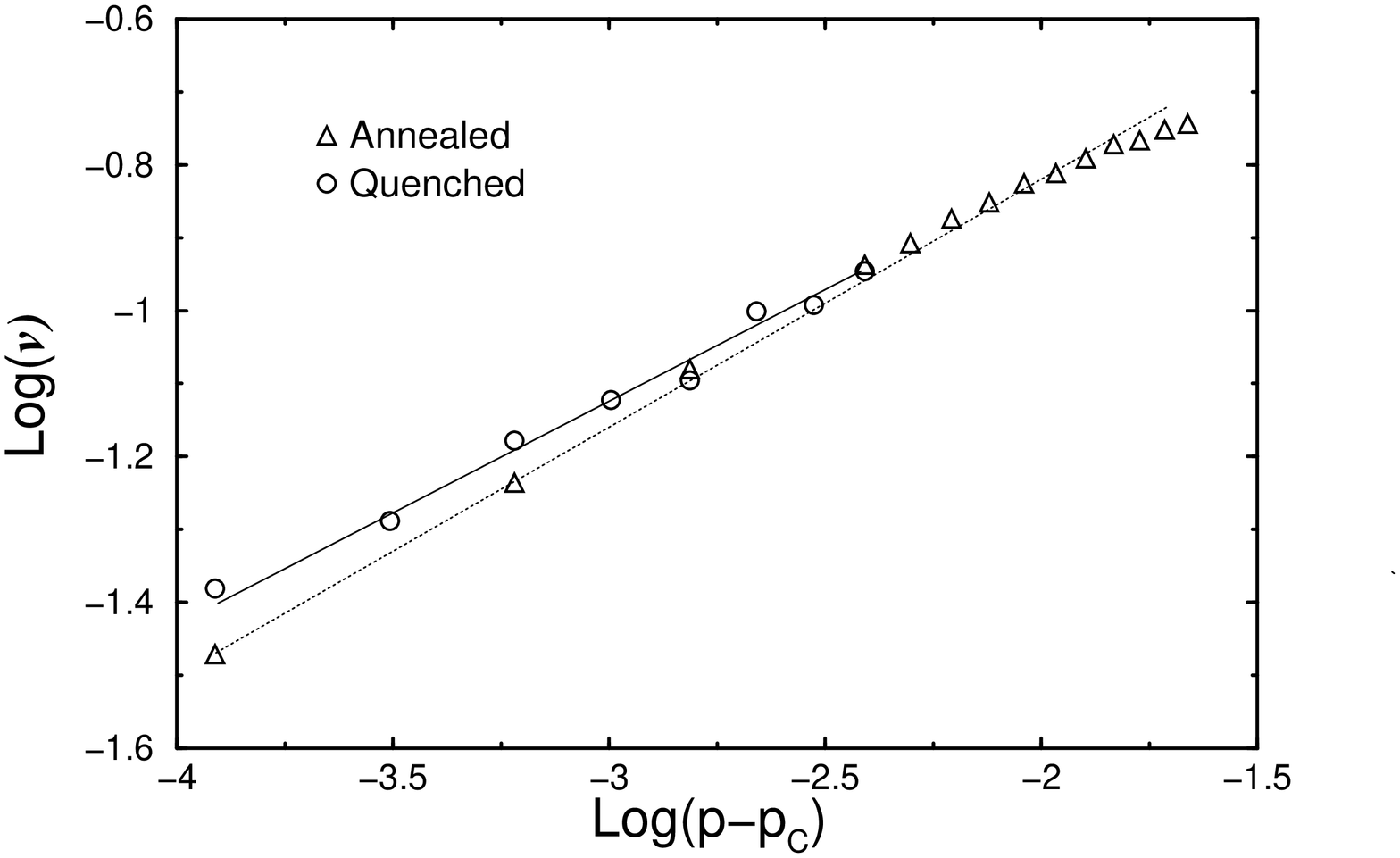}
\caption{Log-log plot of $v$ vs $(p-p_{c})$ for the data from Fig. \ref {f6} for both the quenched and the annealed version on HBSC at $\xi=0.90$ inside a 100$\times$100 lattice .}\label{f8}
\end{figure} 

Taking advantage of the strange man's high efficiency to provoke damage, we have performed several numerical calculations in order to evaluate the propagation speed of the damage $v$ by measuring the time it takes to touch the system boundaries both on the square lattice and on the HBSC. The propagation speed on square lattices of size $L\times L$ was calculated for both the short-range case ($q=0$) and the small-world case ($q\neq0$) in their quenched version, where for the  short-range case we applied an extrapolation technique in order to reduce finite-size effects. This extrapolation was achieved by analysing how the propagation speed of damage $v$ depends on the reciprocal of the lattice size ($1/L$) when one takes the limit $L\rightarrow\infty$ for every value of $p$ considered. Fig. \ref{fig5} shows the propagation speeds for the short-range case for $L$=100, 300 , 700 and $L=\infty$ (extrapolated data) averaged over 100 different runs for each parameter $p$,  while Fig. \ref{fig6} shows the propagation speeds for the small-world case at $q=0.05$ for lattice sizes $L$=100, $L$=200 and $L$=300 averaged over 1000 different runs. It is somewhat surprising to observe in Fig. \ref{fig6} a considerable increase of the propagation speed compared to the case plotted in Fig. \ref{fig5} as well as a clear dependence of speed on the system size. One can observe an almost constant propagation speed above $p=0.45$. 
In other words, it makes little difference for the system dynamics if one is at $p=0.45$ or at $p=0.50$. We have built clusters with boundary-saturation $\xi=0.90$ inside square lattices of size 100$\times$100 and made simulations over 1000 different HBSCs both for the quenched version and the annealed version. Fig. \ref{f6} shows our results for the propagation speed on HBSC at values above and a little below the critical point. The major reason why the propagation speed does not vanish at all for values of $p$ a little below the critical point is the existence of finite-size effects.  Moreover it is interesting to observe that for the annealed version one has a higher propagation speed compared to the quenched version. 

To estimate the critical exponents $\alpha_{q}$ of the speed for the square lattice, we have made a log-log plot of the data from Figs. \ref{fig5} and \ref{fig6} by taking $p_{c}=0.31$ as critical point for the short-range case ($q=0.0$) and  $p_{c}=0.15$ as critical point for the small-world case ($q=0.05$). Fig. \ref{fig7} shows the log-log plot of $v$ vs $(p-p_{c})$ for the extrapolated data from Fig. \ref{fig5}, our estimate for the critical exponent of the speed at ($q=0.0$) is $\alpha_{0}\approx 0.27$, while in Fig. \ref{fig8} a collapse of the data from Fig. \ref{fig6} is shown by plotting the log-log plot of $v/L$ vs $(p-p_{c})$, our estimate for the critical exponent of the speed at ($q=0.05$) is $\alpha_{005}\approx 0.62$. In addition, we also calculated the critical exponent $\alpha$ of the speed on the HBSC at $\xi=0.90$ through a log-log plot of the data from Fig. \ref{f6} as shown in Fig. \ref{f8}. Our results for the critical exponents $\alpha$ of the speed on the HBSC are nearly the same for the quenched and the annealed version. For the quenched version we obtain $\alpha\approx 0.30$ whereas for the quenched version we obtain $\alpha\approx 0.33$ where we use $p_{c}=0.36$ and  $p_{c}=0.26$ as their respective critical points.
\begin{figure}[h]
\includegraphics*[scale=0.40,angle=-90]{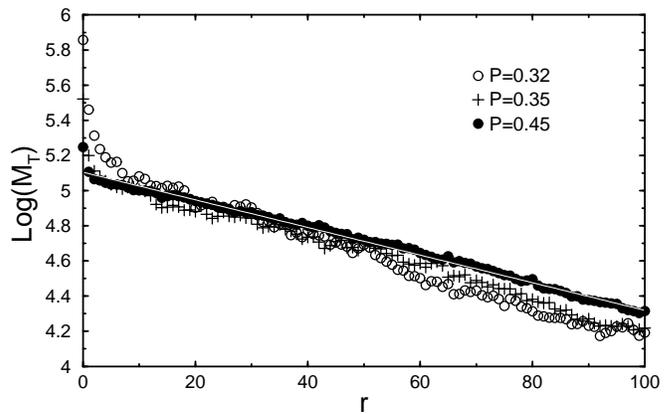}
\caption{Semi-log plot of $M_{T}$ vs the  distance $r$ from the strange man on the square lattice for three different values of $p$ above the critical point, namely $p$= 0.32, 0.35 and 0.45.}\label{fig9}
\end{figure}

\begin{figure*} 
\begin{minipage} [!h]{0.48\linewidth}
\includegraphics*[scale=0.40,angle=-90]{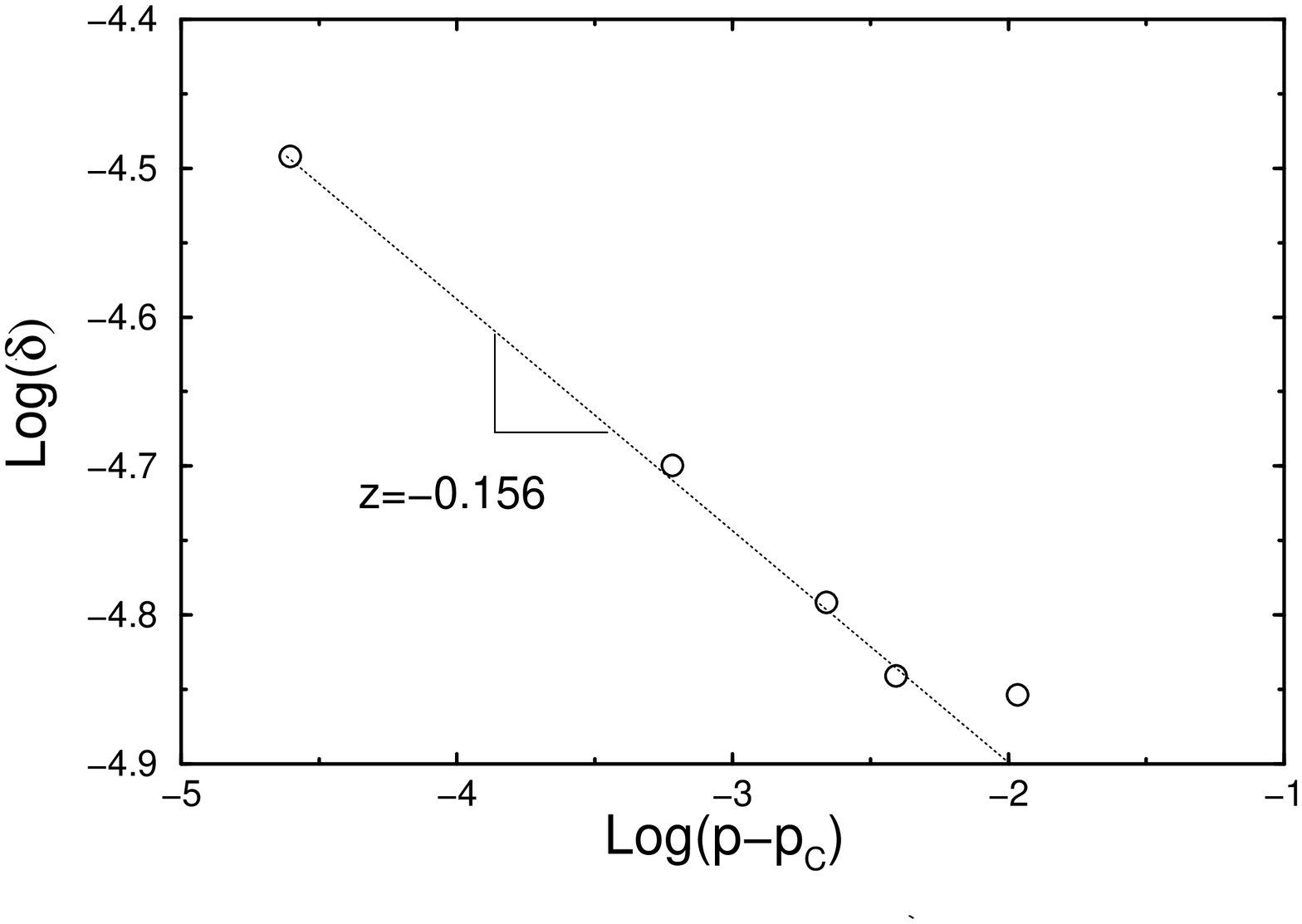}
\caption{Log-log plot of $\delta$ vs $(p-p_{c})$ on a 200$\times$200 lattice. The slope gives the dynamical exponent $z$.}\label{fig10}
\end{minipage}\hfill
 \begin{minipage}[!l]{0.48\linewidth}
 \includegraphics*[scale=0.40,angle=-90]{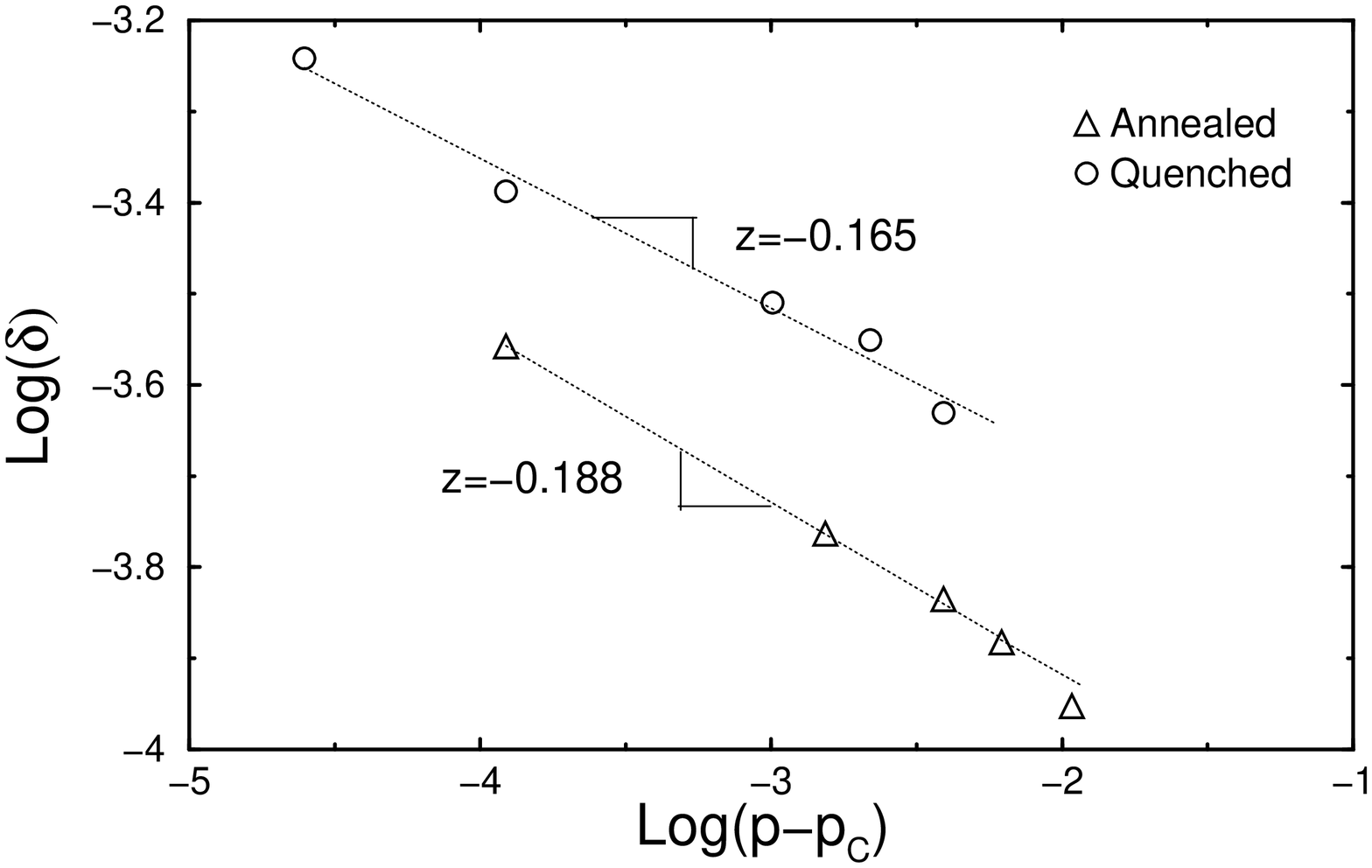}
 \caption{Log-log plot of $\delta$ vs $(p-p_{c})$ on the HBSC inside a 100$\times$100 lattice for both 
 quenched and annealed versions.}\label{fig11}
 \end{minipage}
\end{figure*} 

But how does the damage cloud emitted by the strange man evolve with time when one takes into account different values of $p$? To answer this question, we have performed  simulations in order to determine the dynamical exponent $z$. First one determines the radial distribution of the mean total damage $M_{T}$. This is defined as the mean number of times that the sites along a straight line starting from the strange man have been flipped during the whole time development after the transient time.  That was done by means of a vectorial storage of the total damage for each site along a straight line starting from the strange man on 200$\times$200 lattices and on HBSCs built inside 100$\times$100 lattices over 1000 different configurations for different values of $p$ above the critical point.  On the HBSC both quenched and annealed versions were considered with their respective critical points. Moreover in order to get rid of the transient regime we iterated the systems until about twice their respective touching times for each value of $p$.  Fig. \ref{fig9} shows the semi-log plot of $M_{T}$ against the distance $r$ from the strange man for three different values of $p$ above of the critical point on a 200$\times$200 lattice. One can define the inverse characteristic length $\delta$ for each value of $p$ above the critical point by fitting $M_{T}$ to
\begin{equation} \label{eq:3}
M_{T}(r)= A\exp(\delta r-\phi),
\end{equation}
where $A$ and $\phi$ are constants. 
The dynamical exponent $z$ is defined by 
\begin{equation}
\delta \propto (p-p_{c})^{z},
\end{equation}
where $\delta$ is the inverse characteristic length and $p_{c}$ is the critical point. 
One can determine the dynamical exponent $z$ by a log-log plot of $\delta$ vs $(p-p_{c})$ as shown in Figs. \ref{fig10} and \ref{fig11} for six different values above the critical point on the 200$\times$200 lattice and on the HBSC at $\xi=0.90$ inside a 100$\times$100 lattice, respectively. Our estimate for the dynamical exponent $z$ on the square lattice is $z\approx-0.156$, whereas on the HBSC we find 
\begin{align}
z^{\text{ quenched}}&\approx -0.165\\ z^{\text{annealed}}&\approx -0.188.
\end{align}

The negative sign of $z$ above means that damage decreases as one gets farther from the critical point $p_{c}$. 

\section{Summary}
We have reported some results from simulations to study the damage propagation in the Kauffman model on different graphs such as square lattices and invasion percolation clusters  using a damaging agent that changes its rule all the time. In this paper, we have looked for critical points  by analysing the order parameter for each case. We have not observed any changes at the critical points for both the short-range and small-world case on the square lattice when one introduces small initial damage on the lattice. However the propagation speed is increased due to the strange man's power to damage the system. This has decreased our computational effort because less time is necessary  to calculate the propagation speed as well as the radial distribution of the mean total damage on these graphs. The critical exponents for the propagation speed as well as the dynamical exponents have been calculated with reasonable precision for both the short-range case on the square lattice and on invasion percolation clusters.

We found for the short-range case on the square lattice a critical exponent of the speed $\alpha_{0}\approx0.27$ and a dynamic exponent $z\approx -0.156$. For the small-world case on the square lattice we obtain $\alpha_{005}\approx0.62$. On invasion percolation clusters we obtain critical exponents of the speed $\alpha \approx 0.30$ for the quenched version and $\alpha \approx 0.33$ for the annealed version and  dynamical exponents $z^{\text{ quenched}}\approx -0.165$ and $z^{\text{annealed}}\approx -0.188$. Further we observe a  high sensitivity with respect to damage propagation when one cuts randomly some connections on the square lattice or when one considers invasion percolation clusters with high boundary-saturation.

In addition it would also be interesting to look how the strange man perturbs the basins of attraction in the quenched Kauffman model on the graphs studied here. This analysis is important to better understand the dynamics of the system. That will be the subject of future investigations. 

\section{Acknowledgements} 
 We thank CAPES, CNPq , FUNCAP and the Max Planck prize for financial support.

\end{document}